\documentclass[twocolumn, aps, showpacs]{revtex4}

\usepackage{graphicx,epsfig}
\usepackage{dcolumn}     
\usepackage{bm, rotating}         
\input{Qcircuit}

\begin{document}

\title{Universal fault tolerant quantum computation on bilinear nearest neighbor arrays}

\author{Ashley M. Stephens$^{1,}\footnote{electronic address: a.stephens@physics.unimelb.edu.au}$, Austin G. Fowler$^{2}$ and Lloyd C. L. Hollenberg$^{1}$}

\affiliation{
$^{1}$Centre for Quantum Computer Technology, School of Physics\\
University of Melbourne, Victoria 3010, Australia.\\
$^{2}$Institute for Quantum Computing\\
University of Waterloo, Ontario N2L 3G1, Canada.\\}
\date{\today}

\begin{abstract}
Assuming an array that consists of two parallel lines of qubits and that permits only nearest neighbor interactions, we construct physical and logical circuitry to enable universal fault tolerant quantum computation under the $[[7,1,3]]$ quantum code. A rigorous lower bound to the fault tolerant threshold for this array is determined in a number of physical settings. Adversarial memory errors, two-qubit gate errors, and readout errors are included in our analysis. In the setting where the physical memory failure rate is equal to one-tenth of the physical gate error rate, the physical readout error rate is equal to the physical gate error rate, and the duration of physical readout is ten times the duration of a physical gate, we obtain a lower bound to the asymptotic threshold of $1.96\times10^{-6}$.
\end{abstract}


\maketitle

\section{Introduction}

Quantum computation will potentially enable the efficient solution of computationally difficult problems such as factorization \cite{Shor1}, database searching \cite{Grover1} and quantum system simulation \cite{Aspuru-Guzik1}. However, for any quantum computer system to be feasible it must permit the scalable implementation of fault tolerant quantum error correction (FTEC) \cite{Shor3, Knill3, Gottesman1, Preskill2}. FTEC will exact a reduction in the rate of effective gate failure with each recursive level of encoding provided that the underlying physical failure rate is below some threshold \cite{Kitaev1, Aharonov2}. The specific value of this threshold is influenced by the chosen quantum code, and the efficiency of circuits devised to implement this code, and by the spatial and operational limitations of the physical qubit array. 

Given the considerable interest in quantifying the experimental requirements associated with large-scale quantum computation, a great number of numeric and analytic threshold estimates have been undertaken \cite{seeAliferis1}. Typically, threshold estimates have permitted unlimited range interaction between arbitrary pairs of qubits. In reality, controlled long-range interactions of this type will be difficult to achieve. As many quantum computing proposals are based upon short-range interactions, a more valid assumption is that only nearest neighbor (NN) interactions are available \cite{seeFowler1}. The prevalence of linear nearest neighbor (LNN) proposals also suggests that a restriction may apply to the dimensionality of the qubit array in practice.

Though a threshold has been shown to exist for a one-dimensional array with next nearest neighbor interactions \cite{Gottesman4}, the viability of a LNN system remains an open question. The primary difficulty associated with a strict LNN array is that a single fault in one of the SWAP gates used to interleave adjacent logical qubits prior to a transversal interaction can impart more than one error to a single logical qubit, violating the requirements of FTEC under a distance-3 quantum code. One obvious remedy for this problem would be to implement a larger distance quantum code, such as the $[[25,1,5]]$ Bacon-Shor code \cite{Aliferis2}, the $[[23,1,7]]$ Golay code \cite{Steane4}, or the [[11,1,5]] code \cite{Gottesman5}, which is the most compact known distance-$5$ code. In addition, the use of a LNN array does not preclude the implementation of a large quantum algorithm \cite{Fowler3}. However, due to the increase in the number of qubits and the complexity of the circuitry required to detect and correct multiple-qubit errors, it is expected that such a restriction will incur a significant threshold penalty \cite{Stephens1}. Other ideas under investigation include a two-dimensional NN array, for which a favorable fault tolerant threshold has been obtained \cite{Svore1}, or systems with a mechanism for qubit transport \cite{Kielpinski1, Taylor1}, though there is no clear path to the near-term fabrication of any such architecture. 

In this paper we consider a quasi one-dimensional array that consists of two parallel lines of qubits and permits only NN interactions. A bilinear architecture of this kind appears to be more tractable than any two-dimensional architecture and does not prohibit universal fault tolerant quantum computation under a distance-3 quantum code. Such an architecture could be realized, for example, in a donor-based system, as was originally proposed in \cite{Hollenberg1}, or using superconducting technology \cite{Fowler2}. We note that a CNOT threshold has been estimated for a NN array that increases in length \emph{and} width with each level of concatenation \cite{Szkopek1}. In this paper we present physical and logical circuitry to achieve universal fault tolerant quantum computation on a more constrained bilinear array. In addition, we determine a rigorous lower bound to the threshold for universal fault tolerant computation on this array in a number of physical settings. In the setting where the physical memory failure rate is negligible, we obtain a lower bound to the threshold that is an order of magnitude higher than that presented in \cite{Szkopek1}.

The paper is organized as follows. In Section \ref{FTEC} we review the chosen FTEC protocol and the method for threshold estimation. In Section \ref{circuitry} we describe the adaptation of non-local circuitry for the fault tolerant implementation of a universal set of quantum gates to physical and logical circuitry on a bilinear NN array. Section \ref{thresholds} contains the calculation and presentation of threshold conditions for this array. Section \ref{conclusion} concludes with a summary of results and a description of further work.

\section{Fault tolerant error correction}
\label{FTEC}

The seven qubit $[[7,1,3]]$ code \cite{Steane2} is sufficient to protect against an arbitrary single qubit error and against multiple errors with non-zero probability of success and is an attractive code for concatenation as a number of fault tolerant logical gates can be implemented transversally. For example, as for any CSS code, a logical CNOT is realized by the parallel application of a CNOT from each qubit in the control logical block to each corresponding qubit in the target logical block. The set of logical single qubit gates $\{X,Z,H,S\}$ can also be implemented in this fashion. This property leads to a reduction in the complexity of a number of fault tolerant gates relative to other CSS codes and to non-CSS codes where teleportation may be required. Furthermore, though resource requirements per logical qubit are slightly higher under the $[[7,1,3]]$ code than under a block code \cite{Poulin1}, under the $[[7,1,3]]$ code it is possible to operate on each logical qubit simultaneously, potentially enabling faster computation.

The most gate- and time-efficient method of syndrome extraction under the $[[7,1,3]]$ code is the \emph{encoded block} method \cite{Steane6}. This method requires a seven qubit ancilla block to be prepared in the $[[7,1,3]]$ encoded state $\vert0_L\rangle$. The sequential application of transversal CNOT gates between the ancilla block and the data block allows the extraction of $Z$- and $X$-type syndrome information required for FTEC. For example, the $X$-type syndrome is extracted by applying a transversal CNOT with the ancilla block as the control and the data block as the target. This gate will copy any $Z$ error on the data block to the corresponding location on the ancilla block. Measurement of the ancilla block followed by a classical parity check will determine the location of a single $Z$ error. 

If the circuit for ancilla preparation is not fault tolerant, a single fault during preparation of the state $\vert0_L\rangle$ may lead to more than one error being copied to the data block. For example, during $X$-type syndrome extraction any $X$ errors in the ancilla block will be copied to the data block. While a probabilistic procedure of ancilla verification is typically employed to restore fault tolerance, an alternative procedure has been proposed in which ancilla verification is replaced by a decoding circuit that is applied after the transversal CNOT between the data and ancilla blocks \cite{DiVincenzo3}. Though a single fault during ancilla preparation may still lead to a multiple-qubit error being copied to the data block, the decoding circuit is designed such that it is possible to determine the type and location of this error. A suitable recovery operation will take the form of a product of single qubit $X$ and $Z$ operations. FTEC using ancilla decoding is attractive as the qubits and circuitry that were required to verify the ancilla are no longer necessary. The circuit to implement $X$-type syndrome extraction using the encoded block method and ancilla decoding is shown in Fig$.$ \ref{figure:xtype}.

\begin{figure}
\[ \Qcircuit @C=1em @R=1em @!R {
      \lstick{\vert\psi_L^\prime\rangle} & \qw & \targ & \qw & \gate{R} & \rstick{\vert\psi_L\rangle} \qw \\
      \lstick{\vert{0}\rangle} & \gate{encode} & \ctrl{-1} & \gate{decode} & \meter \cwx[-1]
}\]
\caption{$X$-type syndrome extraction using the encoded block method and ancilla decoding instead of verification. $R$ is a recovery operation that may involve single qubit $X$ and $Z$ operations. }
\label{figure:xtype}
\end{figure}
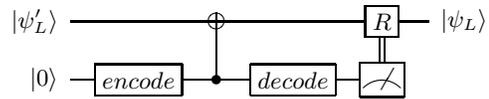

\begin{figure}
\begin{picture}(149,54)
\put(0,19){\line(1,0){10}}
\put(0,35){\line(1,0){10}}
\put(10,15){\framebox(24,25){\footnotesize 0-Ga}}
\put(34,19){\line(1,0){10}}
\put(34,35){\line(1,0){10}}
\put(43,21){\makebox(30,12){$\longrightarrow$}}
\put(71,12){\line(1,0){10}}
\put(71,42){\line(1,0){10}}
\put(81,0){\framebox(24,54){1-Ga}}
\put(105,12){\line(1,0){10}}
\put(105,42){\line(1,0){10}}
\put(115,0){\framebox(24,24){1-EC}}
\put(139,12){\line(1,0){10}}
\put(115,30){\framebox(24,24){1-EC}}
\put(139,42){\line(1,0){10}}
\end{picture}
\caption{To generate an encoded circuit each physical location (0-Ga) is replaced by a rectangle comprising the corresponding logical location (1-Ga) followed by error correction (1-EC) of each logical qubit.}
\label{figure:rec}
\end{figure}
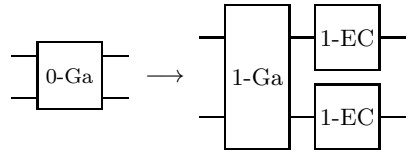

\begin{figure}[t!]
\begin{picture}(112,54)
\put(0,12){\line(1,0){10}}
\put(0,42){\line(1,0){10}}
\put(10,0){\framebox(24,24){1-EC}}
\put(34,12){\line(1,0){10}}
\put(10,30){\framebox(24,24){1-EC}}
\put(34,42){\line(1,0){10}}
\put(44,0){\framebox(24,54){1-Ga}}
\put(68,12){\line(1,0){10}}
\put(68,42){\line(1,0){10}}
\put(78,0){\framebox(24,24){1-EC}}
\put(102,12){\line(1,0){10}}
\put(78,30){\framebox(24,24){1-EC}}
\put(102,42){\line(1,0){10}}
\end{picture}
\caption{Two or more faults must occur within any extended rectangle for an encoded circuit to fail.}
\label{figure:exrec}
\end{figure}
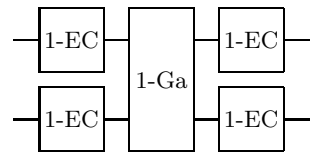

For an architecture that can interact arbitrary pairs of qubits, to generate an encoded quantum circuit each physical qubit is replaced by a logical qubit. Each location in the original circuit (gate, memory, readout or state preparation) is also replaced by a \emph{rectangle} \cite{Aliferis1}. Each rectangle comprises a number of physical locations and executes a fault tolerant logical operation equivalent to the location type being replaced, followed by a FTEC cycle applied to each logical qubit involved in the location. This replacement procedure is shown schematically for a two qubit gate in Fig$.$ \ref{figure:rec}. The resultant encoded circuit will have a higher effective fidelity than the original physical circuit provided that the failure rate of the physical locations is below some threshold. To increase the fidelity of the encoded circuit, this replacement procedure can be applied recursively, whereby the new physical locations are themselves replaced by rectangles. This recursive encoding forms a concatenated quantum code. Note that some technical details of this method need to be modified for the array considered in this paper.

To estimate the fault tolerant threshold an \emph{extended rectangle} is considered  \cite{Aliferis1}, which is identical to a rectangle but is preceded by a FTEC cycle applied to each logical qubit. An extended rectangle is shown schematically in Fig$.$ \ref{figure:exrec}. Under the $[[7,1,3]]$ code, assuming that all circuits are fault tolerant, a single fault in any extended rectangle will not lead to the failure of an encoded circuit as the resultant error or errors will be correctable. Two or more faults must occur within any extended rectangle for a circuit to fail. Assuming that every combination of more than one fault will lead to failure, a lower bound to the threshold is obtained by counting the number of each type of location within the circuit of each extended rectangle, as described in detail in Section \ref{thresholds}.

\section{Physical and logical circuitry}
\label{circuitry}

\begin{figure}
\begin{center}
\resizebox{85mm}{!}{\includegraphics{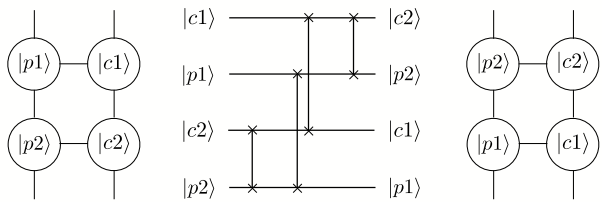}}
\end{center}
\vspace*{-10pt}
\caption{The single-error SWAP of computational qubits $\vert{c1}\rangle$ and $\vert{c2}\rangle$ on a vertical bilinear array. A physical SWAP gate is only ever applied between a computational qubit and a placeholder qubit.}
\label{figure:placeholder}
\end{figure}

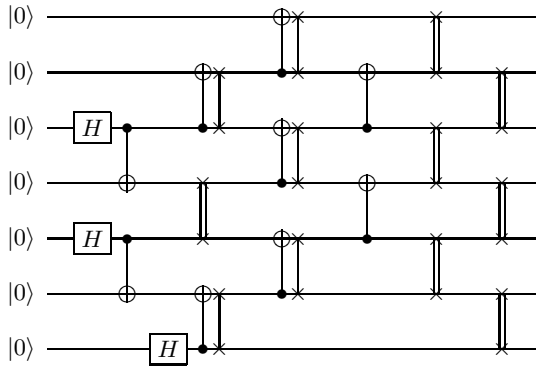
\begin{figure}
\[ \Qcircuit @C=0.27em @R=1em @!R {
      \lstick{\vert0\rangle} & \qw & \qw & \qw & \qw & \qw & \qw & \qw & \qw & \qw & \qw & \qw & \qw & \qw & \qw & \qw & \qw & \targ & \qswap{1} & \qw & \qw & \qw & \qw & \qw & \qw & \qw & \qw & \qw & \qw & \qw & \qw & \qw & \qw & \qw & \qw & \qw & \qswap & \qw & \qw & \qw & \qw & \qw & \qw & \qw & \qw & \qw & \qw & \qw & \qw & \qw & \qw & \qw \\
      \lstick{\vert0\rangle} & \qw & \qw & \qw & \qw & \qw & \qw & \qw & \targ & \qswap & \qw & \qw & \qw & \qw & \qw & \qw & \qw & \ctrl{-1} & \qswap \qwx{-1} & \qw & \qw & \qw & \qw & \qw & \qw & \qw & \qw & \targ & \qw & \qw & \qw & \qw & \qw & \qw & \qw & \qw & \qswap \cwx{-1} & \qw & \qw & \qw & \qw & \qw & \qw & \qw & \qw & \qw & \qswap & \qw & \qw & \qw & \qw & \qw  \\
      \lstick{\vert0\rangle} & \qw & \qw & \qw & \gate{H} & \ctrl{1} & \qw & \qw & \ctrl{-1} & \qswap \qwx{-1} & \qw & \qw & \qw & \qw & \qw & \qw & \qw & \targ & \qswap & \qw & \qw & \qw & \qw & \qw & \qw & \qw & \qw & \ctrl{-1} & \qw & \qw & \qw & \qw & \qw & \qw & \qw & \qw & \qswap & \qw& \qw & \qw & \qw & \qw & \qw  & \qw & \qw & \qw & \qswap \cwx{-1} & \qw & \qw & \qw & \qw & \qw  \\
      \lstick{\vert0\rangle} & \qw & \qw & \qw & \qw & \targ & \qw & \qw & \qswap & \qw & \qw & \qw & \qw & \qw & \qw & \qw & \qw & \ctrl{-1} & \qswap \qwx{-1} & \qw & \qw & \qw & \qw & \qw & \qw & \qw & \qw & \targ & \qw & \qw & \qw & \qw & \qw & \qw & \qw & \qw & \qswap \cwx{-1} & \qw & \qw & \qw & \qw & \qw & \qw & \qw & \qw & \qw & \qswap & \qw & \qw & \qw & \qw & \qw  \\
      \lstick{\vert0\rangle} & \qw & \qw & \qw & \gate{H} & \ctrl{1} & \qw & \qw & \qswap \cwx{-1} & \qw & \qw & \qw & \qw & \qw & \qw & \qw & \qw & \targ & \qswap & \qw & \qw & \qw & \qw & \qw & \qw & \qw & \qw & \ctrl{-1} & \qw & \qw & \qw & \qw & \qw & \qw & \qw & \qw & \qswap & \qw & \qw & \qw & \qw & \qw & \qw & \qw & \qw & \qw & \qswap \cwx{-1} & \qw & \qw & \qw & \qw & \qw  \\
      \lstick{\vert0\rangle} & \qw & \qw & \qw & \qw & \targ & \qw & \qw & \targ & \qswap & \qw & \qw & \qw & \qw & \qw & \qw & \qw & \ctrl{-1} & \qswap \qwx{-1} & \qw & \qw & \qw & \qw & \qw & \qw & \qw & \qw & \qw & \qw & \qw & \qw & \qw & \qw & \qw & \qw & \qw & \qswap \cwx{-1} & \qw & \qw & \qw & \qw & \qw & \qw & \qw & \qw & \qw & \qswap & \qw & \qw & \qw & \qw & \qw  \\
      \lstick{\vert0\rangle} & \qw & \qw & \qw & \qw & \qw & \qw & \gate{H} & \ctrl{-1} & \qswap \qwx{-1} & \qw & \qw & \qw & \qw & \qw & \qw & \qw & \qw & \qw & \qw & \qw & \qw & \qw & \qw & \qw & \qw & \qw & \qw & \qw & \qw & \qw & \qw & \qw & \qw & \qw & \qw & \qw & \qw & \qw & \qw & \qw & \qw & \qw & \qw & \qw & \qw & \qswap \cwx{-1} & \qw & \qw & \qw & \qw & \qw 
      }\]
\caption{Logical circuit for ancilla encoding. Compound gates are used and a double-barred SWAP symbol indicates a single-error SWAP.}
\label{figure:encode}
\end{figure}

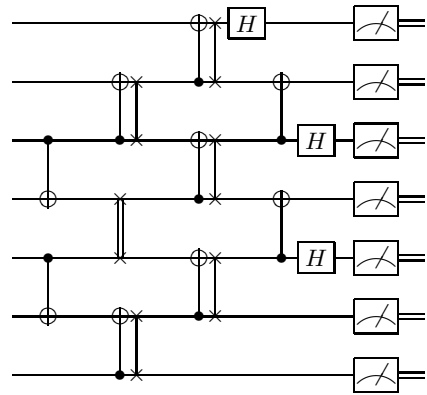
\begin{figure}[t!]
\[ \Qcircuit @C=0.27em @R=1em @!R {
      & \qw & \qw & \qw & \qw & \qw & \qw & \qw & \qw & \qw & \qw & \qw & \qw & \qw & \qw & \qw & \qw & \qw & \qw & \qw & \qw & \targ & \qswap{1} & \qw & \gate{H} & \qw & \qw & \qw & \qw & \meter & \cw & \cw & \cw & \cw  \\
      & \qw & \qw & \qw & \qw & \qw & \qw & \qw & \qw & \qw & \qw & \qw & \targ & \qswap & \qw & \qw & \qw & \qw & \qw & \qw & \qw & \ctrl{-1} & \qswap \qwx{-1} & \qw & \qw & \targ & \qw & \qw & \qw & \meter & \cw & \cw & \cw & \cw  \\
      & \qw & \qw & \qw & \ctrl{1} & \qw & \qw & \qw & \qw & \qw & \qw & \qw & \ctrl{-1} & \qswap \qwx{-1} & \qw & \qw & \qw & \qw & \qw & \qw & \qw & \targ & \qswap & \qw & \qw & \ctrl{-1} & \gate{H} & \qw & \qw & \meter & \cw & \cw & \cw & \cw  \\
      & \qw & \qw & \qw & \targ & \qw & \qw & \qw & \qw & \qw & \qw & \qw & \qswap & \qw & \qw & \qw & \qw & \qw & \qw & \qw & \qw & \ctrl{-1} & \qswap \qwx{-1} & \qw & \qw & \targ & \qw & \qw & \qw & \meter & \cw & \cw & \cw & \cw  \\
      & \qw & \qw & \qw & \ctrl{1} & \qw & \qw & \qw & \qw & \qw & \qw & \qw & \qswap \cwx{-1} & \qw & \qw & \qw & \qw & \qw & \qw & \qw & \qw & \targ & \qswap & \qw & \qw & \ctrl{-1} & \gate{H} & \qw & \qw & \meter & \cw & \cw & \cw & \cw  \\
      & \qw & \qw & \qw & \targ & \qw & \qw & \qw & \qw & \qw & \qw & \qw & \targ & \qswap & \qw & \qw & \qw & \qw & \qw & \qw & \qw & \ctrl{-1} & \qswap \qwx{-1} & \qw & \qw & \qw & \qw & \qw & \qw & \meter & \cw & \cw & \cw & \cw  \\
      & \qw & \qw & \qw & \qw & \qw & \qw & \qw & \qw & \qw & \qw & \qw & \ctrl{-1} & \qswap \qwx{-1} & \qw & \qw & \qw & \qw & \qw & \qw & \qw & \qw & \qw & \qw & \qw & \qw & \qw & \qw & \qw & \meter & \cw & \cw & \cw & \cw  \\
      }\]
\caption{Logical circuit for ancilla decoding. All measurements are made in the computational basis.}
\label{figure:decode}
\end{figure}

\begin{figure}
\[ \Qcircuit @C=2.2em @R=1.9em @!R {
	\lstick{\vert{d1}\rangle} & \qw & \qw & \rstick{\vert{d1}\rangle} \qw \\
	\lstick{\vert{d2}\rangle} & \qw & \qswap & \rstick{\vert{a1}\rangle} \qw \\
	\lstick{\vert{d3}\rangle} & \qswap & \qswap \cwx{-1} &  \rstick{\vert{d2}\rangle} \qw \\
	\lstick{\vert{a1}\rangle} & \qswap \cwx{-1} 	& \qswap &  \rstick{\vert{a2}\rangle} \qw \\
	\lstick{\vert{a2}\rangle} & \qw & \qswap \cwx{-1}&  \rstick{\vert{d3}\rangle} \qw \\
	\lstick{\vert{a3}\rangle} & \qw& \qw & \rstick{\vert{a3}\rangle} \qw \\ 
	}\]
\caption{Logical mesh circuit. The corresponding unmesh circuit is the circuit reversed.}
\label{figure:mesh}
\end{figure}
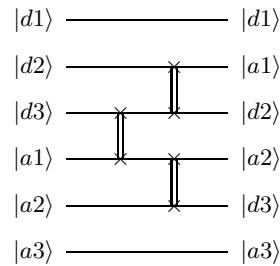

\begin{sidewaysfigure*}
\begin{center}
\resizebox{230mm}{!}{\includegraphics{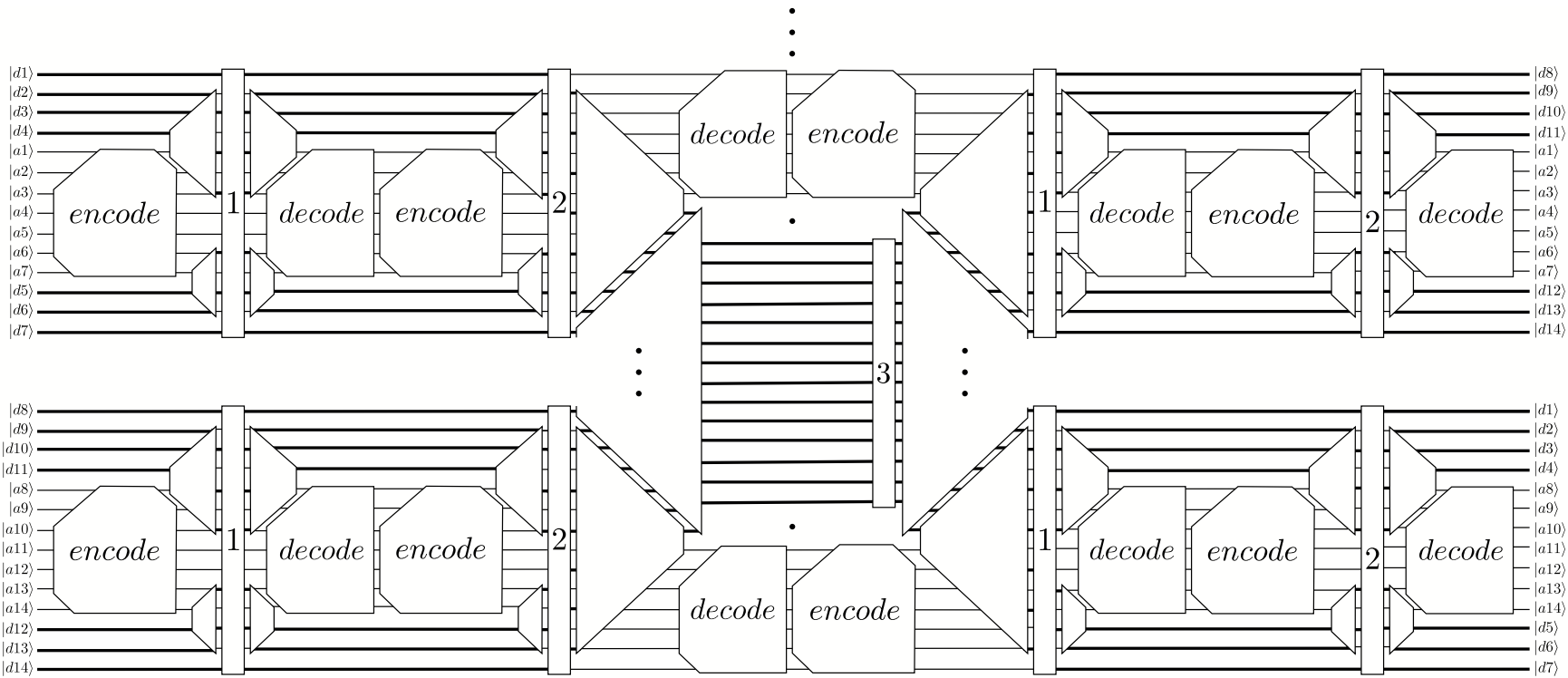}}
\end{center}
\caption{Logical circuit for the SWAP extended rectangle. Bold lines represent data qubits and regular lines represent ancilla qubits. Encode and decode components refer to Fig$.$ \ref{figure:encode} and Fig$.$ \ref{figure:decode} respectively. Triangular components are appropriately sized mesh and unmesh circuits similar to Fig$.$ \ref{figure:mesh}. Rectangular components represent various transversal interactions: 1 is a CNOT with each ancilla qubit the control and each data qubit the target; 2 is a CNOT with each data qubit the control and each ancilla qubit the target, and with $H$ applied to each ancilla qubit before and after the CNOT; and 3 is a transversal SWAP between logical data blocks. Additional SWAP gates are included in transversal interactions where required. Vertical dots indicate the presence of qubits that are not required during the CNOT extended rectangle, but are required to achieve the fault tolerant $T$. The central mesh circuitry is not shown explicitly but all locations are included in the data presented in Table \ref{table:two}.}
\label{figure:lnnSWAP}
\end{sidewaysfigure*}

Adaptation of the chosen FTEC protocol to a bilinear NN qubit array involves the formulation of appropriate quantum circuits for ancilla encoding and for syndrome extraction. To achieve universal computation, circuits are also required for the fault tolerant implementation of a universal set of logical gates. One such set  comprises $\{I, H, X, Z, S, S^{\dagger}, T, \rm{CNOT}\}$. As the [[7,1,3]] code permits the transversal implementation of each of these gates except for $T$, and assuming that single and two qubit gates can be compounded, the circuits required for the error corrected universal set can be adequately represented by $\{I, T, \rm{SWAP}\}$, where $I$ will be referred to as a \emph{memory} location. This is possible because the circuit for the SWAP extended rectangle comprises a similar (and slightly larger) gate layout as that for the CNOT extended rectangle. Therefore, to estimate the threshold for universal quantum computation, the construction of circuitry describing the extended rectangles of memory, SWAP, $T$ and readout is sufficient. Note that the readout extended rectangle includes transversal readout, qubit resetting and logical state preparation.

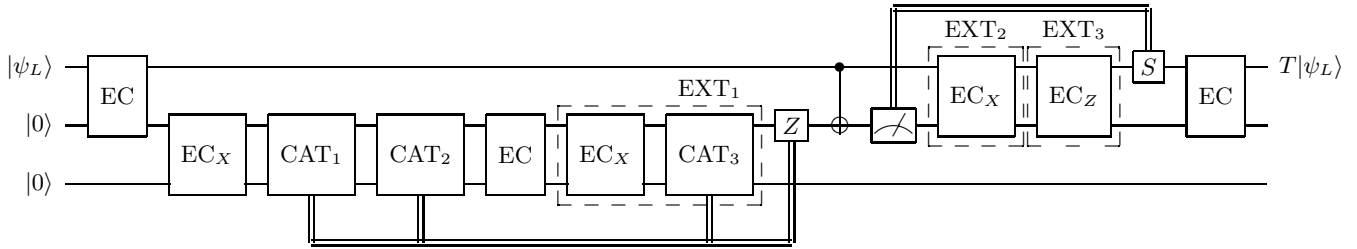
\begin{figure*}
\[ \Qcircuit @C=0.9em @R=1em @!R {
	& & & & & & & & & & & \dstick{\textrm{EXT$_2$}} \cw & \dstick{\textrm{EXT$_3$}} \cw & \cw \\
	\lstick{\vert\psi_L\rangle} & \multigate{1}{\textrm{EC}} & \qw & \qw & \qw & \qw & \qw & \dstick{\textrm{EXT$_1$}} \qw & \qw & \ctrl{1} & \qw \cwx & \multigate{1}{\textrm{EC}_X} & \multigate{1}{\textrm{EC}_Z} & \gate{S} \cwx{-1} & \multigate{1}{\textrm{EC}} & \rstick{T\vert\psi_L\rangle} \qw \\
         \lstick{\vert{0}\rangle} & \ghost{\textrm{EC}} & \multigate{1}{\textrm{EC}_X} & \multigate{1}{\textrm{CAT}_1} & \multigate{1}{\textrm{CAT}_2} & \multigate{1}{\textrm{EC}} & \multigate{1}{\textrm{EC}_X} & \multigate{1}{\textrm{CAT}_3} & \gate{Z} & \targ & \meter \cwx{-1} & \ghost{\textrm{EC}_X} & \ghost{\textrm{EC}_Z} & \qw & \ghost{\textrm{EC}} & \qw \\
         \lstick{\vert{0}\rangle} & \qw & \ghost{\textrm{EC}_Z} & \ghost{\textrm{CAT}_1} & \ghost{\textrm{CAT}_2} & \ghost{\textrm{EC}} & \ghost{\textrm{EC}_Z} & \ghost{\textrm{CAT}_3} & \cwx \qw & \qw & \qw & \qw & \qw & \qw & \qw & \qw \gategroup{3}{7}{4}{8}{.7em}{--} \gategroup{2}{12}{3}{12}{.7em}{--} \gategroup{2}{13}{3}{13}{.7em}{--} \\
       & & & \cwx & \cw \cwx & \cw & \cw& \cw \cwx & \cw \cwx & & & & & 
}\]
\caption{Fault tolerant circuit for the $T$ extended rectangle that is based on the implementation in Fig$.$ \ref{figure:T1}. EC is a full error correction cycle and EC$_X$ and EC$_Z$ are partial error correction cycles that extract the $X$- and $Z$-type syndrome respectively, as shown in Fig$.$ \ref{figure:xtype}. Each CAT refers to the circuit in Fig$.$ \ref{figure:T2}. The preparation of the state $\vert{A}_\frac{\pi}{4}\rangle=T\vert + \rangle$ is as follows: EC$_X$ will prepare the state $\vert0_L\rangle$ from the state $\vert0\rangle$, then two measurements of the operator $SX=TXT^\dagger$ are performed followed by a full error correction cycle. If an error is detected in the EC following CAT$_2$ or in either of CAT$_1$ or CAT$_2$, or if the measurement results of CAT$_1$ and CAT$_2$ disagree, then the ancilla preparation is aborted, the ancilla is reset and the additional circuitry denoted by EXT$_1$ is used. To complete the preparation, logical Z is applied to the ancilla block conditional upon the eigenvalue of the operator $SX=TXT^\dagger$ being $-1$. Following the interaction between the ancilla and data, logical ${S}$ is applied to the data block conditional upon the associated measurement outcome being $1$. If an error is detected by the second syndrome measurement within the first EC, and if EXT$_1$ has not already been applied, the additional circuitry denoted by EXT$_2$ is used. If an error is detected in EXT$_2$ the additional circuitry denoted by EXT$_3$ is used.}
\label{figure:Tmain}
\end{figure*}

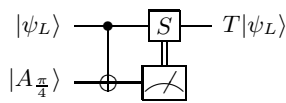
\begin{figure}
\[ \Qcircuit @C=1em @R=1em {
      \lstick{\vert\psi_L\rangle} & \ctrl{1} & \gate{S} & \rstick{T \vert\psi_L\rangle} \qw \\
      \lstick{\vert{A}_\frac{\pi}{4}\rangle} & \targ & \meter \cwx[-1]
}\]
\caption{Implementation of the logical rotation $T=exp(\frac{-i \pi Z}{8})$, where $\vert{A}_\frac{\pi}{4}\rangle=T\vert + \rangle$. Logical ${S}$ is applied to the data block conditional upon the measurement outcome being $1$. The required state $\vert{A}_\frac{\pi}{4}\rangle$ is prepared by the circuit in Fig$.$ \ref{figure:T2}.}
\label{figure:T1}
\end{figure}

In the absence of any other fault tolerant transport mechanism, for a NN architecture to be feasible it must include a fault tolerant SWAP mechanism at the physical level and at each logical level. Under the $[[7,1,3]]$ code, one fault during this process should impart no more than one error to any logical qubit. To satisfy this requirement, the qubit array is divided between \emph{computational} qubits, which include all data and ancilla qubits, and \emph{placeholder} qubits, the states of which are inconsequential to any computational processes \cite{Gottesman4}. Then, as shown in Fig$.$ \ref{figure:placeholder}, the \emph{single-error} SWAP of two computational qubits is realized through a sequence of physical SWAP gates that move two computational qubits past each other on the bilinear array. During this sequence each physical SWAP is only ever applied between a computational qubit and a placeholder qubit. As the states of placeholder qubits are inconsequential, a single faulty SWAP will impart only one error to a single computational qubit and, therefore, only one error to a single logical qubit. This forms what is effectively a SWAP-based fault tolerant transport mechanism. A single-error logical SWAP can be constructed from this underlying mechanism in order to satisfy the fault tolerant requirement at all levels. At any level above the physical level, the bilinear NN physical qubit array becomes what is effectively a LNN logical qubit array with the added property of single-error logical SWAP gates. Thus, in the formulation of circuitry describing the required extended rectangles, a distinction is made between \emph{logical circuitry} and \emph{physical circuitry}:

\begin{figure}
\[ \Qcircuit @C=1em @R=1em @!R {
         \lstick{\vert0_L\rangle} & \qw & \gate{T^\dagger} & \targ & \gate{T} & \qw & \gate{Z} & \rstick{\vert{A}_\frac{\pi}{4}\rangle} \qw \\
         \lstick{\vert{0}\rangle} & \gate{cat} & \qw & \ctrl{-1} & \qw & \gate{uncat} & \meter \cwx[-1]
}\]
\caption{Circuit to prepare the required state $\vert{A}_\frac{\pi}{4}\rangle=T\vert + \rangle$. Logical ${Z}$ is applied to the data block conditional upon the measured eigenvalue of the operator $SX=TXT^\dagger$ being $-1$, though in the circuit in Fig$.$ \ref{figure:Tmain} this is only after the eigenvalue has been determined after two or three separate measurements. The cat state decoding procedure suggested in \cite{DiVincenzo3} is used to detect (but not correct) errors. Also note that a $T^{\dagger}T$ pair can be eliminated when this circuit is repeated in the circuit in Fig$.$ \ref{figure:Tmain}.}
\label{figure:T2}
\end{figure}
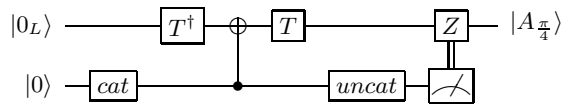

\emph{Logical circuitry.} With the introduction of single-error SWAP gates, LNN circuitry can be made fault tolerant using a distance-$3$ code at any logical level. Logical circuitry to implement FTEC simply follows from the equivalent non-local circuitry. Figures \ref{figure:encode} and \ref{figure:decode} show the [[7,1,3]] ancilla encoding and decoding circuits respectively where all circuitry is valid for any level above the physical level and where a single-error logical SWAP is indicated by a double-barred SWAP symbol. Additional LNN \emph{mesh} and \emph{unmesh} circuits, shown in Fig$.$ \ref{figure:mesh}, are used to interleave and then separate data and ancilla blocks prior to and after $X$- and $Z$-type syndrome extraction and to interleave adjacent logical blocks prior to any inter-logical transversal interaction. Circuitry for the SWAP extended rectangle is constructed from these basic components and is shown in Fig$.$ \ref{figure:lnnSWAP}. Construction of the memory, $T$ and readout extended rectangles similarly follows from non-local circuitry. For the $T$ extended rectangle, our starting point was the non-local fault tolerant circuit presented in \cite{Aliferis1} which we then made more compact, reducing both the qubit and gate counts, the details of which can be found in Figs$.$ \ref{figure:Tmain}-\ref{figure:T2}. Finally, for the readout extended rectangle, the state $\vert 0_L\rangle$ is prepared from the state $\vert0\rangle$ by the circuit in Fig$.$ \ref{figure:xtype}.

We note that a number of measurement outcomes will result in a larger $T$ circuit, as shown in Fig$.$ \ref{figure:Tmain}. For example, an additional partial or full FTEC cycle is required prior to the classically controlled $S$. This is to prevent the combination of an $X$ and a $Z$ error on different qubits in a single logical block, which may arise due to a single fault during syndrome extraction, from transforming to a combination of a $Y$ error and an $X$ or $Z$ error on different qubits. To be consistent with the rigorous assumptions made in Section \ref{thresholds}, and as our analysis requires that the depth of the $T$ rectangle does not dependent on particular measurement outcomes, we consider the largest circuit that can result from a single fault.

\begin{figure}[t!]
\begin{center}
\resizebox{80mm}{!}{\includegraphics{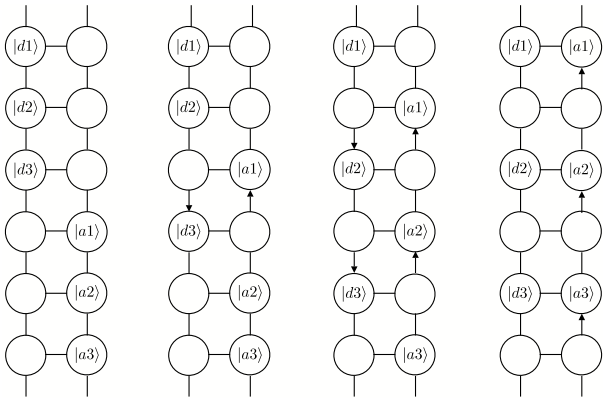}}
\end{center}
\vspace*{-10pt}
\caption{An example of a physical mesh circuit that is more efficient than the circuit generated by simply replacing each SWAP in the LNN circuit in Fig$.$ \ref{figure:mesh} with the circuit shown in Fig$.$ \ref{figure:placeholder}. Four discrete steps are shown.}
\label{figure:physical}
\end{figure}

\emph{Physical circuitry.}
To construct physical level circuitry from LNN circuitry it would be sufficient to replace every single-error logical  SWAP by the single-error SWAP shown in Fig$.$ \ref{figure:placeholder}. To obtain more gate- and time-efficient circuitry we note that the logical LNN mesh circuit has a much more efficient physical gate implementation, as shown in Fig$.$ \ref{figure:physical}, than that suggested by the naive implementation shown in Fig$.$ \ref{figure:placeholder}. Additional efficiency was achieved by including gates between physical qubits that are oriented perpendicular to the direction of the bilinear array. This resulted in circuitry that is significantly more efficient than that presented in \cite{Szkopek1}. All physical level circuitry was optimized by hand in this manner.

\begin{figure}
\begin{picture}(226,112)
\put(0,8){\line(1,0){28}}
\put(28,0){\framebox(16,16){EC}}
\put(44,8){\line(1,0){28}}
\put(72,0){\framebox(16,16){EC}}
\put(88,8){\line(1,0){28}}
\put(116,0){\framebox(16,16){EC}}
\put(132,8){\line(1,0){28}}
\put(160,0){\framebox(16,16){EC}}
\put(176,8){\line(1,0){28}}
\put(204,0){\framebox(16,16){EC}}
\put(220,8){\line(1,0){6}}
\put(0,32){\line(1,0){6}}
\put(0,56){\line(1,0){6}}
\put(6,24){\framebox(16,40){SW}}
\put(22,32){\line(1,0){6}}
\put(22,56){\line(1,0){6}}
\put(28,24){\framebox(16,16){EC}}
\put(44,32){\line(1,0){6}}
\put(28,48){\framebox(16,16){EC}}
\put(44,56){\line(1,0){6}}
\put(50,24){\framebox(16,40){SW}}
\put(66,32){\line(1,0){6}}
\put(66,56){\line(1,0){6}}
\put(72,24){\framebox(16,16){EC}}
\put(88,32){\line(1,0){6}}
\put(72,48){\framebox(16,16){EC}}
\put(88,56){\line(1,0){6}}
\put(94,24){\framebox(16,40){SW}}
\put(110,32){\line(1,0){6}}
\put(110,56){\line(1,0){6}}
\put(116,24){\framebox(16,16){EC}}
\put(132,32){\line(1,0){6}}
\put(116,48){\framebox(16,16){EC}}
\put(132,56){\line(1,0){6}}
\put(138,24){\framebox(16,40){SW}}
\put(154,32){\line(1,0){6}}
\put(154,56){\line(1,0){6}}
\put(160,24){\framebox(16,16){EC}}
\put(176,32){\line(1,0){6}}
\put(160,48){\framebox(16,16){EC}}
\put(176,56){\line(1,0){6}}
\put(182,24){\framebox(16,40){SW}}
\put(198,32){\line(1,0){6}}
\put(198,56){\line(1,0){6}}
\put(204,24){\framebox(16,16){EC}}
\put(220,32){\line(1,0){6}}
\put(204,48){\framebox(16,16){EC}}
\put(220,56){\line(1,0){6}}
\put(0,80){\line(1,0){6}}
\put(6,72){\framebox(192,16){T}}
\put(198,80){\line(1,0){6}}
\put(204,72){\framebox(16,16){EC}}
\put(220,80){\line(1,0){6}}
\put(0,104){\line(1,0){6}}
\put(6,96){\framebox(60,16){R}}
\put(66,104){\line(1,0){6}}
\put(72,96){\framebox(16,16){EC}}
\put(88,104){\line(1,0){6}}
\put(94,96){\framebox(60,16){R}}
\put(154,104){\line(1,0){6}}
\put(160,96){\framebox(16,16){EC}}
\put(176,104){\line(1,0){28}}
\put(204,96){\framebox(16,16){EC}}
\put(220,104){\line(1,0){6}}
\end{picture}
\caption{Circuitry is arranged at all levels such that the duration of one $T$ rectangle is equal to that of five memory rectangles, the duration of one readout rectangle is equal to that of two memory rectangles and the duration of one SWAP rectangle is equal to that of one memory rectangle.}
\label{figure:synch}
\end{figure}
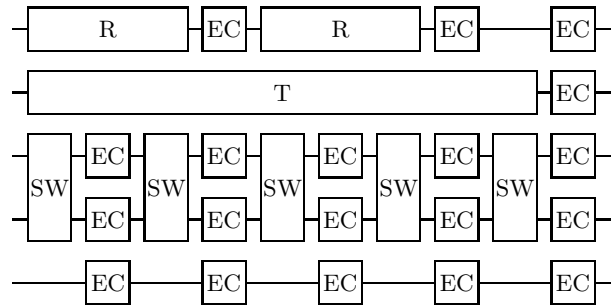

To facilitate analysis, at the physical level and at all logical levels the duration of rectangles describing all locations were arranged to permit parallel locations, as illustrated in Fig$.$ \ref{figure:synch}. Furthermore, since, in principle, a $T$ could occur anywhere in a circuit at any time, during encoding every physical qubit is replaced by 21 physical qubits. It is expected that the detailed arrangement of gates and qubits will ultimately, however, be an automated process. This process may involve the dense packing of FTEC cycles during memory and may also incorporate routing to avoid any defective regions that are identified during some initial characterization of the physical architecture.\\

\section{Thresholds}
\label{thresholds}

To obtain our threshold estimate it was assumed that faults would occur with a probability specific to the physical location type and that this probability was equal for each location of the same type. It was also assumed that every combination of two or more faults would lead to a combination of errors that was not correctable under the $[[7,1,3]]$ code. This assumption ensures a rigorous lower bound to the threshold. Note that it is possible to find and neglect pairs of faults that lead to correctable errors, thereby relaxing this pessimistic assumption and enabling a more precise determination of the threshold. This has been demonstrated in previous work to raise the lower bound by a factor of between 3.0 and 4.5 \cite{Aliferis1, Aliferis2}. Given these assumptions, the failure probability of any physical circuit is
\begin{equation}
\begin{array}{lccl}
      & 1 & - &(1-p_{m})^{N_{m}}(1-p_{S})^{N_{S}}(1-p_{r})^{N_{r}} \\
      &   & - & N_{m} p_{m}(1-p_{m})^{N_{m}-1}(1-p_{S})^{N_{S}}(1-p_{r})^{N_{r}} \\
      &   & - & (1-p_{m})^{N_{m}}N_{S} p_{S}(1-p_{S})^{N_{S}-1}(1-p_{r})^{N_{r}} \\
      &   & - & (1-p_{m})^{N_{m}}(1-p_{S})^{N_{S}}N_{r} p_{r}(1-p_{r})^{N_{r}-1},         
\end{array}
\end{equation}
where $p_{m}$, $p_{S}$ and $p_{r}$ are the failure probabilities of physical memory, gate and readout locations respectively, and where $N_{m}$, $N_{S}$ and $N_{r}$ are the numbers of physical memory, gate and readout locations in the circuit. The second term corresponds to the failure of no locations. The third and fourth terms correspond to the failure of a single memory location and a single two-qubit gate location respectively, where applied single- and two-qubit errors are chosen adversarially. The final term corresponds to the failure of a single readout location, where a bit flip is applied to the classical result. Similarly, the failure probabilities of the physical level memory, SWAP, $T$ and readout extended rectangles are given by 
\begin{widetext}
\begin{equation}
\begin{array}{lcccl}
p_{1j} & = & 1 & - &(1-p_{0m})^{N_{0mj}}(1-p_{0S})^{N_{0Sj}}(1-p_{0r})^{N_{0rj}} \\
      &   &   & - & N_{0mj} p_{0m}(1-p_{0m})^{N_{0mj}-1}(1-p_{0S})^{N_{0Sj}}(1-p_{0r})^{N_{0rj}} \\
      &   &   & - & (1-p_{0m})^{N_{0mj}}N_{0Sj} p_{0S}(1-p_{0S})^{N_{0Sj}-1}(1-p_{0r})^{N_{0rj}} \\
      &   &   & - & (1-p_{0m})^{N_{0mj}}(1-p_{0S})^{N_{0Sj}}N_{0rj} p_{0r}(1-p_{0r})^{N_{0rj}-1},         
\end{array}
\end{equation}
\end{widetext}
where $p_{0m}$, $p_{0S}$ and $p_{0r}$ are the probabilities of failure of physical memory, two-qubit gate and readout locations respectively and where $N_{0ij}$ is the number of $i$-type locations in the physical level (level-1) $j$ extended rectangle for $i=\{m, S, r\}$ (memory, SWAP, readout) and $j=\{m, S, T, r\}$ (memory, SWAP, $T$, readout). To account for the expected variation in timing of gates and readout, $N_{0mj}$ is a expressed as a function of $t_r$, which is defined as the ratio of physical readout time to physical gate time. Therefore, as the relative duration of physical readout increases the number of memory locations in any extended rectangle will increase proportionately.

The failure probabilities of logical level extended rectangles can be expressed similarly, but as functions of the failure probabilities of the physical level extended rectangles, $p_{1m}$, $p_{1S}$, $p_{1T}$ and $p_{1r}$, as given by Eqs$.$ (2-5), rather than of the physical locations. For example, the failure probability of the first logical level (level-2) $T$ extended rectangle is
\begin{widetext}
\begin{equation}
\begin{array}{lcccl}
p_{2T} & = & 1 & - & (1-p_{1m})^{N_{1mT}}(1-p_{1S})^{N_{1ST}}(1-p_{1T})^{N_{1TT}}(1-p_{1r})^{N_{1rT}} \\
& & & -&N_{1mT} p_{1m}(1-p_{1m})^{N_{1mT}-1}(1-p_{1S})^{N_{1ST}}(1-p_{1T})^{N_{1TT}}(1-p_{1r})^{N_{1rT}} \\
& & & -&(1-p_{1m})^{N_{1mT}}N_{1ST} p_{1S}(1-p_{1S})^{N_{1ST}-1}(1-p_{1T})^{N_{1TT}}(1-p_{1r})^{N_{1rT}} \\
& & & -&(1-p_{1m})^{N_{1mT}}(1-p_{1S})^{N_{1ST}} N_{1TT} p_{1T}(1-p_{1T})^{N_{1TT}-1}(1-p_{1r})^{N_{1rT}} \\
&& & -&(1-p_{1m})^{N_{1mT}}(1-p_{1S})^{N_{1ST}}(1-p_{1T})^{N_{1TT}}N_{1rT} p_{1r}(1-p_{1r})^{N_{1rT}-1},
\end{array}
\end{equation}
\end{widetext}
where $N_{1iT}$ is the number of $i$-type locations in the logical level $T$ extended rectangle for $i=\{m, S, T, r\}$. Polynomials that describe the logical level memory, SWAP and readout extended rectangles take a similar form to Eq$.$ (6), though there are no $T$ gates in any of these circuits. Due to the self-similarity of circuitry at all logical levels, polynomials describing level-$n$ extended rectangles are identical to the corresponding level-2 polynomials, though they are expressed in terms of the failure probabilities of level-($n$$-$$1$) extended rectangles. Furthermore, as the of duration circuitry at all logical levels is artificially set, there is no requirement for time-dependence to be included in any logical level polynomials. Tables \ref{table:one} and \ref{table:two} show values of  $N_{0ij}$ and $N_{1ij}$ determined from the physical and logical circuitry presented in Section \ref{circuitry}. If memory locations are not neglected, the $T$ extended rectangle contains the most locations at the physical level and at all logical levels. As $T$ is required to achieve universality, the threshold for universal fault tolerant computation is given by the $T$ threshold.

\begin{table}
\begin{tabular}{c|c|c|c|c}
 & $i=m$ & $i=S$ & $i=r$ & depth \\
\cline{1-5}
$j=m$ & $654+28t_r$ & $408$ & $40$ & $41+2t_r$ \\
$j=S$ & $1002+56t_r$ & $1122$ & $80$ & $41+2t_r$ \\
$j=T$ & $3032+133t_r$ & $1228$ & $128$ & $205+10t_r$ \\
$j=r$ & $1045+42t_r$ & $510$ & $57$ & $82+4t_r$ 
\end{tabular}
\caption{Number of $i$-type locations in the physical level (level-1) $j$ extended rectangle for $i=\{m, S, r\}$ and $j=\{m, S, T,r\}$. The depth of each rectangle is also shown where the fundamental time scale is given by the duration of a physical gate.}
\label{table:one}
\end{table}
\begin{table}
\begin{tabular}{c|c|c|c|c|c}
 & $i=m$ & $i=S$ & $i=T$ & $i=r$ & depth \\
\cline{1-6}
$j=m$ & $558$ & $204$ & $0$ & $28$ & $38$ \\
$j=S$ & $824$ & $603$ & $0$ & $56$ &$38$ \\
$j=T$ & $2605$ & $619$ & $28$ & $98$ &$190$ \\
$j=r$ & $974$ & $255$ & $0$ & $42$ &$76$
\end{tabular}
\caption{Number of $i$-type locations in the logical level-$n$ $j$ extended rectangle for $i=\{m, S, T, r\}$ and $j=\{m, S, T, r\}$. The depth of each rectangle is also shown where the time scale is given by the duration of a level-($n$$-$$1$) memory rectangle.}
\label{table:two}
\end{table}

\begin{figure}
\begin{center}
\resizebox{85mm}{!}{\includegraphics{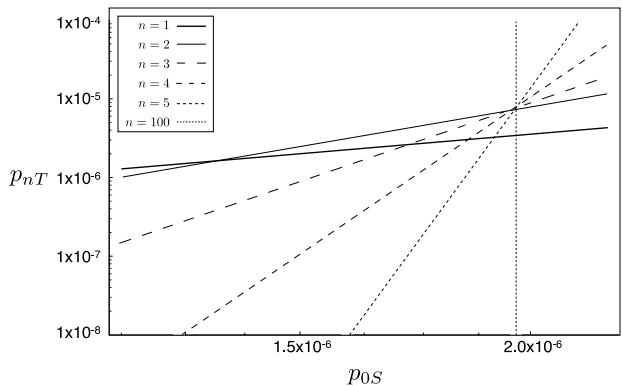}}
\end{center}
\vspace*{-10pt}
\caption{$p_{nT}(p_{0S}, 0.1, 1.0, 10.0)$ for $n=\{1,2,3,4,5,100\}$. Lower bounds to the levels 2-5 and 100 $T$ thresholds are $1.36\times10^{-6}$, $1.72\times10^{-6}$, $1.85\times10^{-6}$, $1.91\times10^{-6}$ and $1.96\times10^{-6}$ respectively.}
\label{figure:plot1}
\end{figure}

Following the ideas presented in \cite{Svore2}, the level-$n$ $T$ threshold is where the probability of failure of the level-$n$ $T$ extended rectangle is equal to the probability of failure of the level-1 $T$ extended rectangle. As any polynomial at any level can be expressed as a function of the failure probabilities of physical locations by the recursive substitution of lower level polynomials, it is possible to determine a lower bound to the level-$n$ threshold by solving
\begin{equation}
p_{nT}(p_{0S}, p_{0m}, p_{0r},t_r)=p_{1T}(p_{0S}, p_{0m}, p_{0r},t_r),
\end{equation}
for $p_{0S}$, $p_{0m}$ and $p_{0r}$. To estimate a physical gate threshold error rate it is necessary to specify the ratios of the physical memory failure rate and the physical readout failure rate to the failure rate of all other physical gates, denoted by $R_m$ and $R_r$ respectively. The specification of $R_m$, $R_r$ and $t_r$ is referred to as a physical \emph{setting}. Therefore, determination of a lower bound to the level-$n$ threshold as an explicit gate failure rate in any physical setting involves solving
\begin{equation}
p_{nT}(p_{0S}, R_m, R_r, t_r )=p_{1T}(p_{0S}, R_m, R_r, t_r ),
\label{condition}
\end{equation}
for $p_{0S}$. Figure \ref{figure:plot1} shows the failure rate of the levels 1-5 and 100 $T$ extended rectangles as a function of $p_{0S}$ in the setting where $R_m=0.1$, $R_r=1.0$ and $t_r=10.0$ and includes a lower bound to each of the levels 2-5 and 100 $T$ thresholds in this setting. 
Levels 2-3 thresholds are particularly useful as they indicate the maximum gate failure rates tolerable in a practical quantum computer. Note that if the gate failure rate is equal to the level-$n$ threshold, if between 2 and $n$$-$$1$ levels of encoding are used the encoded circuit will operate with a lower effective fidelity than at only one level of encoding. Only with levels of encoding beyond $n$, and, therefore, only with greater resources, will a higher fidelity encoded circuit be achieved. Realistically, gate failure rates must be at least an order of magnitude less than these thresholds to obtain significant benefit from FTEC.

To make a comparison with thresholds previously obtained for other qubit arrays it is necessary to consider the asymptotic $T$ threshold. For a physical gate failure rate equal to this threshold recursive encoding maintains the failure rate of the logical $T$ extended rectangle. A lower bound to the asymptotic $T$ threshold is approximated by the level-$100$ $T$ threshold, also included in Fig$.$ \ref{figure:plot1}. For a NN array that increases in length and width with each level of concatenation, in the setting where $R_m=0.0$, $R_r=1.0$ and $t_r=1.0$ a CNOT threshold of $1.20\times10^{-7}$ has been obtained \cite{Szkopek1}. In the identical setting, for a bilinear NN array we obtain a lower bound to the asymptotic $T$ threshold of $2.88\times10^{-6}$. For a two-dimensional NN array, in the setting where $R_m=0.1$, $R_r=1.0$ and $t_r=1.0$, accounting for pairs of faults that lead to correctable errors, a threshold for universal computation of $1.85\times10^{-5}$ was obtained \cite{Svore1}. In the identical setting, for the bilinear NN array we obtain a lower bound to the asymptotic $T$ threshold of $2.05\times10^{-6}$. 

\begin{figure}
\begin{center}
\resizebox{84mm}{!}{\includegraphics{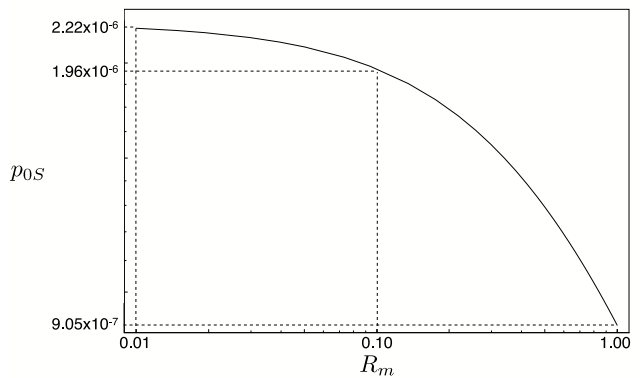}}
\end{center}
\vspace*{-10pt}
\caption{Asymptotic $T$ threshold as a function of $R_m$ in the setting where $R_r=1.0$ and $t_r=10.0$.}
\label{figure:plot2}
\end{figure}

\begin{figure}
\begin{center}
\resizebox{85mm}{!}{\includegraphics{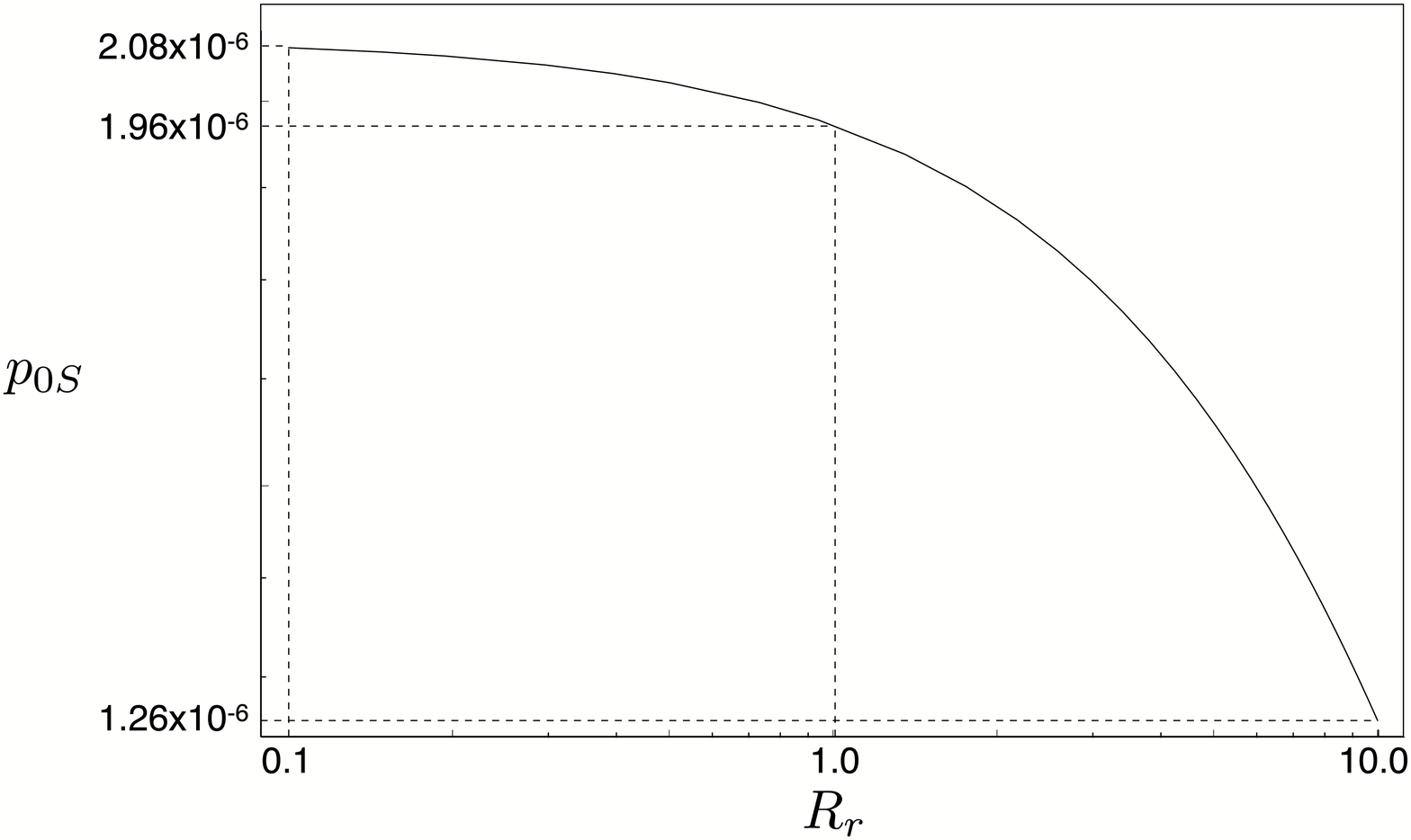}}
\end{center}
\vspace*{-10pt}
\caption{Asymptotic $T$ threshold as a function of $R_r$ in the setting where $R_m=0.1$ and $t_r=10.0$.}
\label{figure:plot3}
\end{figure}

\begin{figure}[t!]
\begin{center}
\resizebox{85mm}{!}{\includegraphics{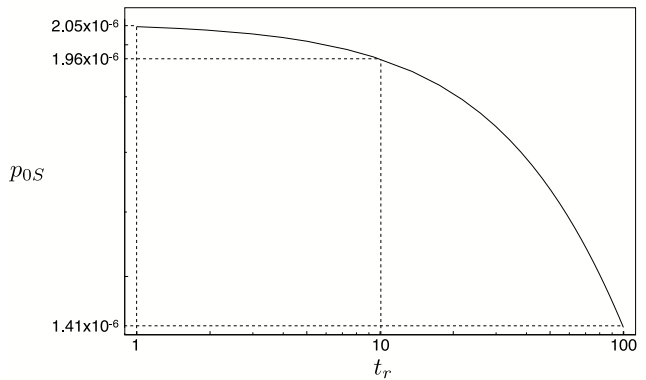}}
\end{center}
\vspace*{-10pt}
\caption{Asymptotic $T$ threshold as a function of $t_r$ in the setting where $R_m=0.1$ and $R_r=1.0$.}
\label{figure:plot4}
\end{figure}

Finally, to observe the dependency of the asymptotic threshold on the physical setting, $R_m$, $R_r$ and $t_r$ were varied from the initial setting where $R_m=0.1$, $R_r=1.0$ and $t_r=10.0$, as shown in Figs$.$ \ref{figure:plot2}-\ref{figure:plot4} respectively. The greatest variation of the threshold is observed when $R_m$ is varied. This is expected as memory locations are the predominant type in most of the physical and logical level extended rectangles. However, the overall variation from the initial threshold is not great. For example, even assuming the extreme parameters $R_m=1.0$, $R_r=100$ and $t_r=1000$, the threshold is reduced by less than two orders of magnitude to $3.78\times10^{-8}$.

\section{Conclusion}
\label{conclusion}

This paper presents physical and logical circuitry to achieve universal fault tolerant quantum computation on a bilinear NN array. The lower bound to the asymptotic $T$ threshold derived from this circuitry is $1.96\times10^{-6}$. This result represents a significant improvement over a CNOT threshold previously obtained for a less constrained array \cite{Szkopek1}. We note that this improvement is despite the requirement of universality and despite the application of a more rigorous method of threshold estimation. As this improvement can be attributed to a reduction in the area of the physical and logical circuitry required to implement FTEC, it is expected that further advances in the efficiency of FTEC protocols and associated circuitry will correspond to further improvements in the threshold.

While the results presented in this paper are highly relevant to existing NN architectures and to architectures that may be restricted to NN interactions between logical qubits \cite{Hollenberg1, Fowler2}, further work is required to determine the threshold for universal fault tolerant computation on a LNN array under one or more larger distance codes. Such a result would help in assessing the viability of the large number of LNN architectures under development. Further work devising specific physical schemes of long-range interaction or transport that are fault tolerant is also desirable to permit the design of architectures with higher thresholds.

AMS is grateful to Zachary Evans and Simon Devitt for helpful discussions and suggestions. AMS and LCLH are supported by the Australian Research Council, the Australian Government, and by the US National Security Agency (NSA), Advanced Research and Development Activity (ARDA), and the Army Research Office (ARO) under contract number W911NF-04-1-0290. 

\vspace{110pt}

\bibliographystyle{unsrt}

\begin{thebibliography}{10}
\bibitem{seeAliferis1}
see \cite{Aliferis1} and references therein. More recent threshold estimates are contained in \cite{DiVincenzo3} and \cite{Aliferis2}.
\end{thebibliography}

\end{document}